# Extremely $^{54}$Cr- and $^{50}$Ti-rich presolar oxide grains in a primitive meteorite: Formation in rare types of supernovae and implications for the astrophysical context of solar system birth.

Larry R. Nittler, Conel M. O'D. Alexander, Nan Liu[1], Jianhua Wang

Department of Terrestrial Magnetism, Carnegie Institution of Washington, 5241 Broad Branch Rd NW, Washington, DC 20015, USA; lnittler@ciw.edu

Short title: Extreme $^{54}$Cr-rich supernova presolar oxide grains

---

[1] Present address: Dept. of Physics, Washington University, 1 Brookings Drive, St. Louis, MO, 63130, USA

Abstract

We report the identification of 19 presolar oxide grains from the Orgueil CI meteorite with substantial enrichments in $^{54}$Cr, with $^{54}$Cr/$^{52}$Cr ratios ranging from 1.2 to 56 times the solar value. The most enriched grains also exhibit enrichments at mass 50, most likely due in part to $^{50}$Ti, but close-to-normal or depleted $^{53}$Cr/$^{52}$Cr ratios. There is a strong inverse relationship between $^{54}$Cr enrichment and grain size; the most extreme grains are all <80 nm in diameter. Comparison of the isotopic data with predictions of nucleosynthesis calculations indicate that these grains most likely originated in either rare, high-density Type Ia supernovae (SNIa), or in electron-capture supernovae (ECSN) which may occur as the end stage of evolution for stars of mass 8−10 $M_\odot$. This is the first evidence for preserved presolar grains from either type of supernova. An ECSN origin is attractive since these likely occur much more frequently than high-density SNIa, and their evolutionary timescales (~20 Myr) are comparable to those of molecular clouds. Self-pollution of the Sun's parent cloud from an ECSN may explain the heterogeneous distribution of n-rich isotopic anomalies in planetary materials, including a recently reported dichotomy in Mo isotopes in the solar system. The stellar origins of three grains with solar $^{54}$Cr/$^{52}$Cr, but anomalies in $^{50}$Cr or $^{53}$Cr, as well as of a grain enriched in $^{57}$Fe, are unclear.

1. Introduction

Meteoritic measurements have revealed that the protosolar nebula from which the Sun and planets formed 4.57 billion years ago was isotopically heterogeneous in several neutron-rich isotopes of iron-peak elements whose primary nucleosynthetic origins are still unclear. These include $^{48}$Ca, $^{50}$Ti, and $^{54}$Cr, all of which have been found to be variable and correlated among bulk planetary materials (e.g., Trinquier et al. 2009; Chen et al. 2011; Dauphas et al. 2014). Meyer et al. (1996) showed that the production of $^{48}$Ca requires high-temperature, low-entropy conditions and suggested that these could be obtained in rare types of high-density Type Ia supernovae (SNIa). Woosley (1997) soon thereafter showed that such environments could indeed be sources of not only $^{48}$Ca, but also the other highly n-rich isotopes. More recently, Wanajo et al. (2013a) suggested an alternative site for synthesis of these species, electron-capture supernovae (ECSN) occurring as the end-stage of the evolution of "super-AGB" stars (of mass ~8−10 $M_\odot$). Wanajo et al. showed that ECSN can produce similarly low-entropy conditions to those of the high-density SNIa considered by Woosley (1997). Whereas the synthesis of $^{48}$Ca requires unusual low-entropy conditions, both $^{50}$Ti and $^{54}$Cr can be made by neutron-capture processes in AGB stars and Type II supernovae (SNII) as well. Nonetheless, co-variations of anomalies in all three isotopes in early solar system materials suggest a common origin.

Prior work by Dauphas et al. (2010) and Qin et al. (2011) revealed the existence of tiny, highly $^{54}$Cr-rich oxide nanoparticles (<100 nm in diameter) in acid residues of the Orgueil (CI) carbonaceous chondrite meteorite. These grains were suggested to be the carriers of the $^{54}$Cr variations observed at bulk meteorite scales and their high $^{54}$Cr/$^{52}$Cr ratios require formation in supernovae. However, the spatial resolution of the NanoSIMS ion probes used in the Dauphas et al. and Qin et al. studies, 400–800 nm, was substantially coarser than the size of the analyzed grains (<100 nm), and the magnitudes of the measured isotopic anomalies were consequently lower limits, precluding the ability to distinguish SNIa from SNII origins. Whereas the previously observed maximum $^{54}$Cr enrichment was ~2.5×solar, simulations suggested that the true enrichments in the grains were up to ~50×solar (Qin et al. 2011). Here we report the use of a new high-resolution O$^-$ primary ion source on the NanoSIMS to accurately determine the Cr isotopic compositions of sub-100-nm grains. We have confirmed the presence of extreme $^{54}$Cr anomalies of up to 57×solar, with correlated anomalies at mass 50, most likely due at least in part to $^{50}$Ti. The measured isotopic compositions of the most extreme grains favor an origin in SNIa or ECSN over SNII, with important implications for the astrophysical environment of the Sun's birth.

## 2. Methods

We analyzed the same acid residue of the Orgueil CI chondrite previously studied by Qin et al. (2011). This sample was prepared by CsF/HCl dissolution of a small amount of bulk Orgueil followed by destruction of organic matter by O plasma ashing and deposition of the residue onto a high-purity Au foil. We used a Cameca NanoSIMS 50L ion microprobe equipped with a Hyperion RF plasma O source (Oregon Physics, LLC) to map Cr isotopes in 195 15×15 µm areas of the sample mount. A 2-pA, ~100-nm-diameter O$^-$ primary ion beam was used in multicollection imaging mode to produce 256×256-pixel positive secondary ion images of the four Cr isotopes as well as $^{48}$Ti (to correct for possible $^{50}$Ti interference on $^{50}$Cr) and $^{56,57}$Fe (both to correct $^{54}$Cr for any $^{54}$Fe interference and to search for Fe-isotope anomalies). Thirty repeated image cycles were acquired for each area, for a total counting time of 32 minutes, or 30 ms per pixel. One area was re-analyzed for $^{48,49}$Ti, $^{51}$V, $^{50,52,54}$Cr, and $^{56}$Fe under similar conditions, but with a raster size of 10×10 µm; this measurement was made to investigate whether an isotopic anomaly detected at mass 50 was attributable to Ti, V, or Cr, as discussed below. Following the NanoSIMS analysis, the analyzed areas were imaged in a JEOL 6500F scanning electron microscope (SEM).

We used our custom L'image software to analyze the NanoSMS images, which were corrected for counting system deadtime and spatial shifts between cycles. Each image contained hundreds of individual sub-µm grains and isotopic ratios were internally normalized to the average values in each image. Anomalous grains were identified both manually by examining isotopic ratio images and by automatic image segmentation. Interference corrections were made by linear fitting of isotopic ratios (e.g., $^{54}$Cr/$^{52}$Cr) versus elemental ratios (e.g., $^{56}$Fe/$^{52}$Cr) as discussed by Qin et al. (2011). Grains were determined to be anomalous if their isotopic ratios differed from normal by more than 4σ. Typical uncertainties for individual ~100-nm grains were 12%, 9%, and 17% for $^{50}$Cr/$^{52}$Cr, $^{53}$Cr/$^{52}$Cr, and $^{54}$Cr/$^{52}$Cr, respectively.

## 3. Results

Out of about 60,000 Cr-rich grains identified in the images, we identified 22 grains with >4σ Cr isotope anomalies (Table 1). Of these, 19 have $^{54}$Cr enrichments ranging from ~1.2 to 57 times solar. The remaining three have normal $^{54}$Cr within errors, but resolved anomalies in $^{50}$Cr or $^{53}$Cr.

One $^{54}$Cr-rich grain, 2_81, also has a moderate $^{57}$Fe anomaly with $^{57}$Fe/$^{56}$Fe 1.4±0.09 times solar. NanoSIMS and SEM images for the most $^{54}$Cr-rich grain, 2_37, are shown in Figure 1.

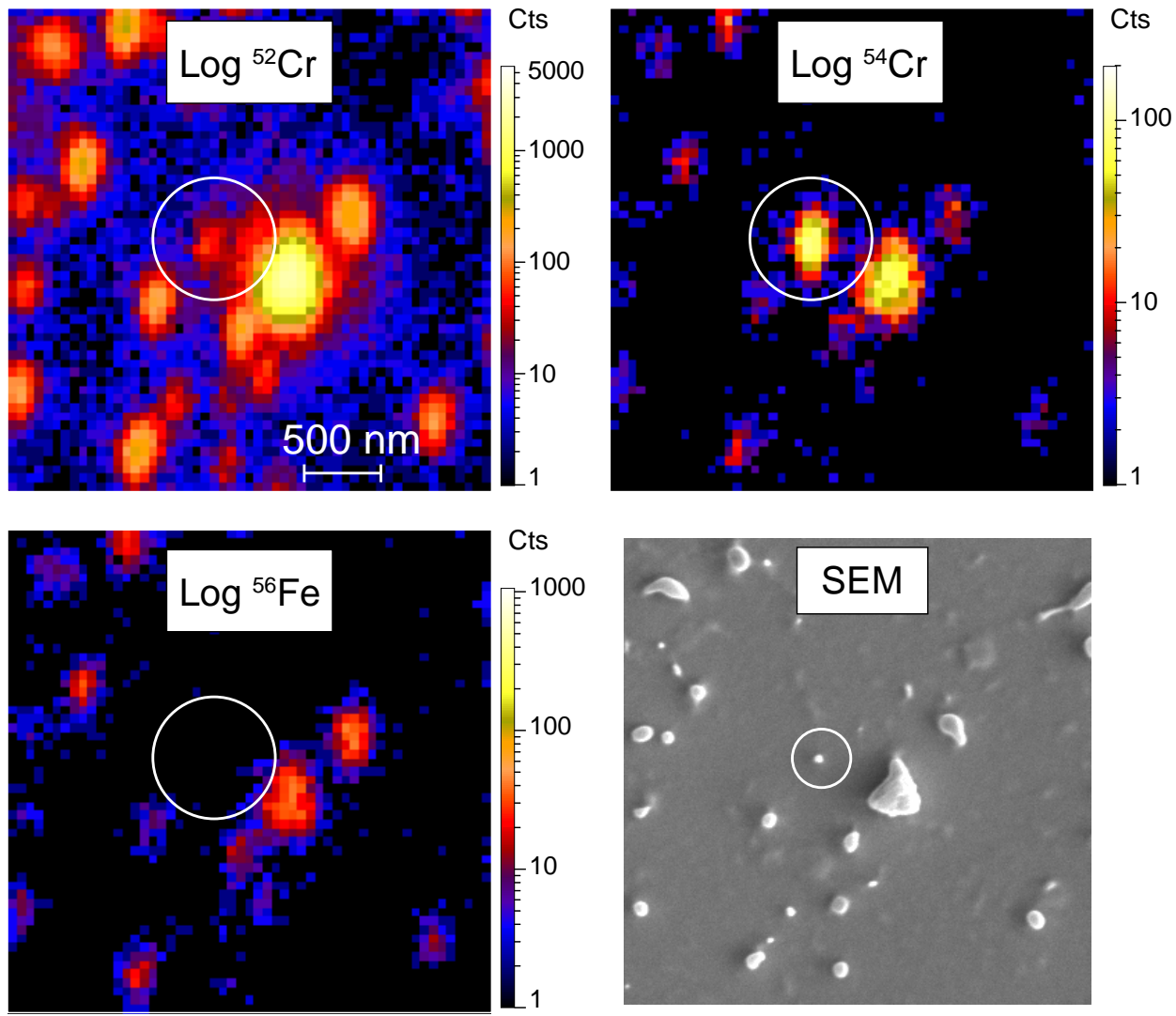

Figure 1. NanoSIMS and SEM images of the most extreme $^{54}$Cr-rich grain, the 60-nm diameter 2_37.

SEM analysis indicated that, after the NanoSIMS measurements, the anomalous grains ranged in diameter from 50 nm to 300 nm (Table 1, Fig. 2). Fifteen grains were completely resolved from other grains on the mount, two were in piles such that we could not determine the identity of the anomalous grain, and the other five were close to other grains that must have contributed some Cr to the isotopic measurements. In Figure 2, we plot the $^{54}$Cr/$^{52}$Cr ratios versus grain diameter for the 14 resolved grains with $^{54}$Cr anomalies. The approximate 4σ detection limit as a function of grain size is indicated by a solid curve. There is a clear inverse relationship between magnitude of

$^{54}$Cr anomaly and grain size, with the most extreme anomalies, >2× solar, only observed in the smallest (<80 nm) grains. This size-dependence is consistent with the results of Dauphas et al. (2010), who found the largest bulk $^{54}$Cr anomaly in the smallest grain size separates of their Orgueil acid residue (see their Fig. 7).

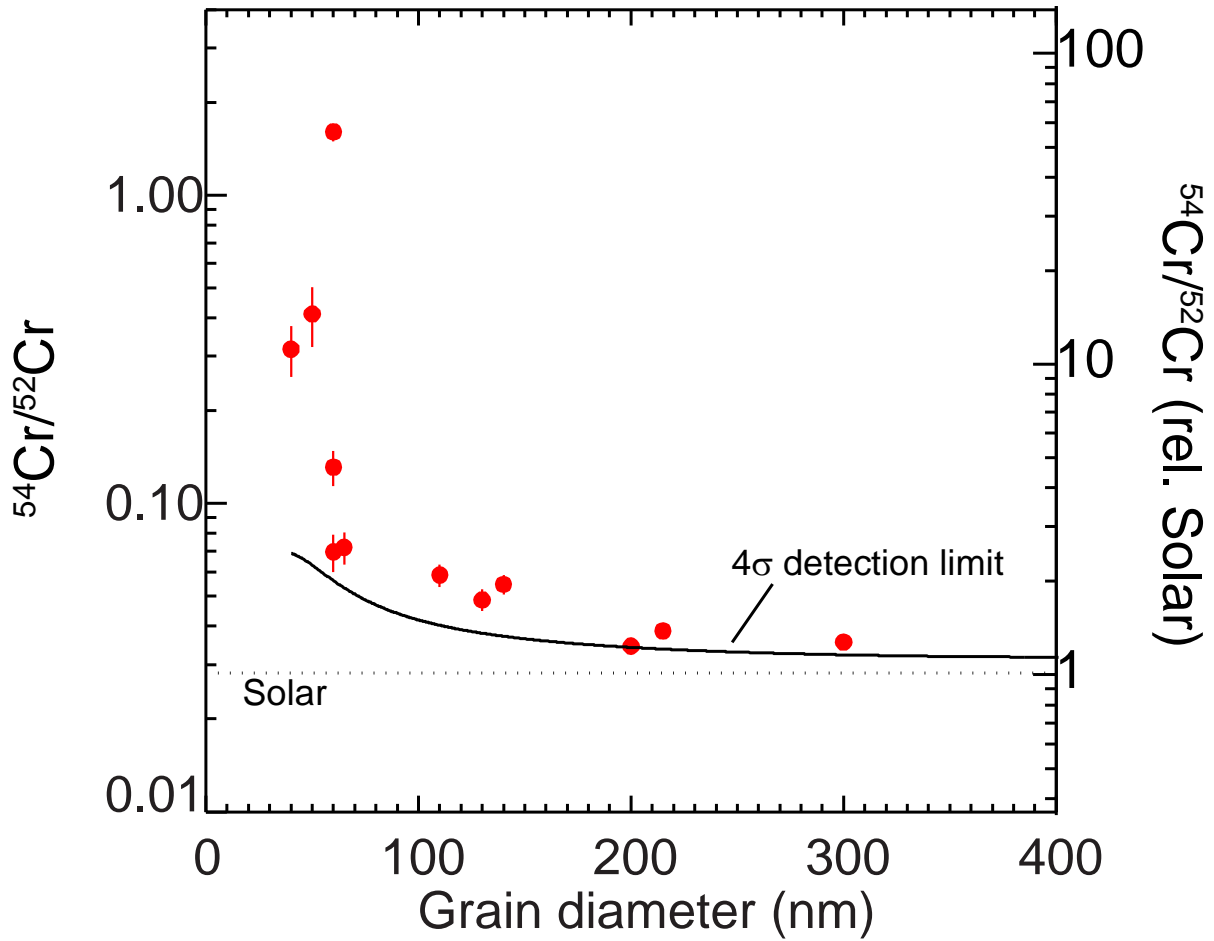

Figure 2. Measured $^{54}$Cr/$^{52}$Cr ratios as a function of grain diameter for fourteen $^{54}$Cr-rich grains for which SEM analysis indicated that they were fully spatially resolved. Black curve indicates approximate 4σ detection limit, the criterion used here to identify a grain as isotopically anomalous. Error bars are 1σ.

The $^{54}$Cr/$^{52}$Cr and $^{53}$Cr/$^{52}$Cr ratios for the anomalous grains are compared with those reported by Dauphas et al. (2010) and Qin et al. (2011) in Figure 3. Lower limits are shown for the seven grains not fully resolved from adjacent grains on the sample mount. The $^{54}$Cr-rich grains mostly have close-to-solar $^{53}$Cr/$^{52}$Cr ratios (Fig. 3), but the most anomalous grain (2_37) also has a large enrichment at mass 50; smaller mass 50 excesses are seen in the other grains with $^{54}$Cr/$^{52}$Cr>0.1 (3.5×solar). It is ambiguous as to whether these anomalies are due to $^{50}$Cr or $^{50}$Ti, as these isotopes were unresolved in the mass spectrometer. For these five grains, we thus calculated two $^{50}$Ti/$^{48}$Ti ratios (Table 1, Figure 4b,c) based on the measured signals at mass 48, 50, and 52 – one by correcting for $^{50}$Cr with the assumption of a solar $^{50}$Cr/$^{52}$Cr ratio and one with the assumption that the signal is due purely to $^{50}$Ti. As a further test, we re-analyzed grain 2_37 for Ti isotopes and $^{51}$V. No V signal was seen, ruling out $^{50}$V as the source of the anomaly for this grain. Titanium counts were extremely low; within the region of interest defined by the $^{54}$Cr signal we detected 1 $^{48}$Ti ion and 3 $^{49}$Ti ions. This would correspond to an extremely anomalous $^{49}$Ti/$^{48}$Ti ratio (3 compared to the solar ratio of 0.07446), strongly supporting that the mass 50 anomaly is due primarily or entirely to Ti for this grain. However, we also identified other pixels in the image, not associated with any grain, with similar counts at mass 49, and thus cannot rule out the $^{49}$Ti signal being an instrumental background.

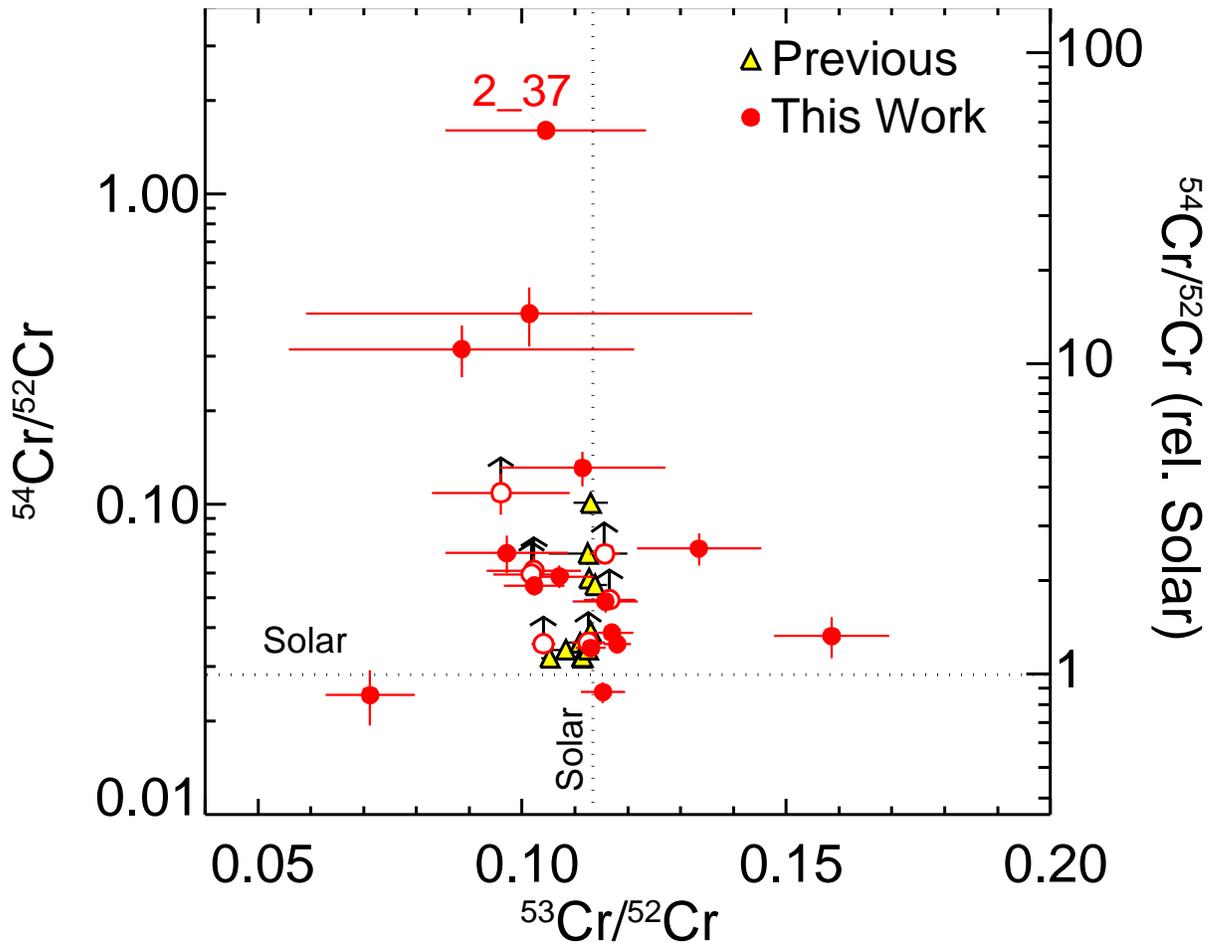

Figure 3: Chromium isotopic ratios in anomalous grains identified here and in previous work (Dauphas et al. 2010; Qin et al. 2011). The higher spatial resolution of the new measurements reveals a larger range of isotopic anomalies, although our (1σ) error bars are larger in the present work due to shorter counting times during NanoSIMS isotopic measurements. Arrows indicate that the measured ratios are lower limits for grains not fully spatially resolved on the NanoSIMS mount (Table 1).

4. Discussion

The high spatial resolution isotopic measurements reported here have confirmed the presence of extreme (>10× solar) $^{54}$Cr enrichments in a few oxide nanoparticles as suggested by Qin et al. (2011). However, most of the new grains span a similar range of $^{54}$Cr/$^{52}$Cr ratios to those seen in the previous lower-resolution work, indicating that the majority of $^{54}$Cr-rich presolar nanoparticles are less extreme. Nonetheless, the average $^{54}$Cr/$^{52}$Cr ratio of the 19 new $^{54}$Cr-rich grains (Table 1) is 0.17, or about 6 times solar, clearly dominated by the most extreme grains. We thus focus our discussion here on these as they will have the largest impact on bulk Cr isotopic variations in planetary materials.

Chromium isotopes are affected by nuclear reactions occurring in asymptotic giant branch (AGB) stars and in supernovae of various types. Neutron-capture reactions in AGB stars are expected to lead to an increase of at most 40% in $^{54}$Cr/$^{52}$Cr (Zinner et al. 2005), and these stars can thus be ruled out as sources of most of the $^{54}$Cr-rich grains reported here. We thus compare the isotope data for the grains to predictions of models of three types of supernovae (Fig. 4): core-collapse SNII from stars more massive than ~10 $M_\odot$ (e.g., Rauscher et al. 2002), high-density SNIa (Woosley 1997), and electron-capture SNe from core collapse of stars of mass ~8-10 $M_\odot$ (Wanajo et al. 2013a).

Chromium-54 is made in massive stars by neutron capture during core He burning and shell C and Ne burning ("weak *s*-process") and ejected when they explode as Type II SNe. Predicted average compositions of the four interior $^{54}$Cr-rich zones of a 15 $M_\odot$ SNII model of Woosley & Heger (2007), labeled by the most abundant elements in each zone (Meyer et al. 1995), are shown on Figure 4, with symbol sizes scaled proportionally to the mass of Cr ejected by each zone. Although these zones can reach very high $^{54}$Cr/$^{52}$Cr ratios as observed in the most extreme grains, the most $^{54}$Cr-rich regions are generally also either highly enriched or highly depleted in $^{53}$Cr, in stark contrast to the grain data, which largely show $^{53}$Cr/$^{52}$Cr ratios close to solar even for the most extreme $^{54}$Cr-rich grains. This result argues strongly against a SNII origin for the most $^{54}$Cr-rich grains. In contrast, many of the less extreme grains have compositions consistent with expectations for the He/C zone of SNII, but it may be difficult to explain oxide minerals forming in such a C-rich environment (C/O>>1).

In contrast to the SNII predictions, models of the other SN types predict strong production of $^{54}$Cr without (slow-conflagration high-density SNIa) or with only a moderate amount of (fast-deflagration SNIa and ECSN) concomitant production of $^{53}$Cr and, as a result, these models are in much better agreement with the grain data in Figure 4a. SNIa are believed to occur due to accretion of material from a companion star onto a white dwarf star in a binary system. As the mass of the white dwarf approaches the Chandrasekhar limit of ~1.44 $M_\odot$, C fusion is ignited leading to an explosion. Most models of SNIa nucleosynthesis do not predict production of n-rich isotopes, but Meyer et al. (1996) suggested that electron capture reactions in high-density SNIa could lead to low-entropy, neutron-rich environments conducive to formation of $^{48}$Ca and other n-rich isotopes. The calculations of Woosley (1997) bear this out; shown on Figure 4 are predictions for high-density SNIa calculated for five different densities and two speeds ("fast" and "slow") of the C deflagration flame front leading to the explosion.

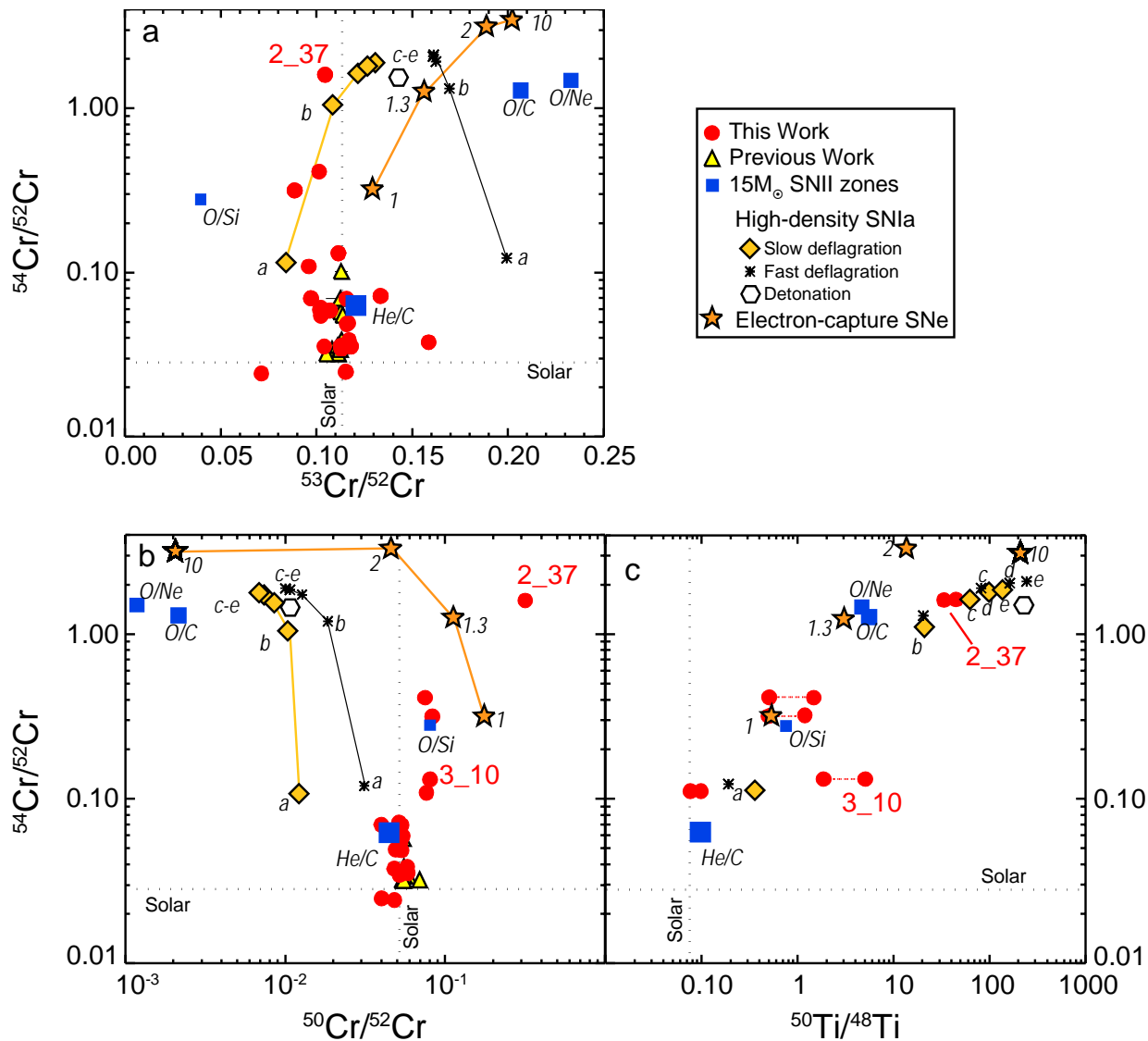

Figure 4. Chromium and Ti isotopic ratios of anomalous grains (circles and triangles) compared with predictions of supernova models. Two $^{50}$Ti/$^{48}$Ti ratios are calculated for each anomalous grain (see text and Table 1), connected by red-dotted lines. Type II supernova predictions (squares) are for the 15 $M_\odot$ model of Woosley & Heger (2007) with zones labeled according to the most abundant isotopes; the symbol sizes are proportional to the total mass of Cr contained in each zone. High-density Type Ia supernovae predictions (diamonds, asterisks and hexagons) from Woosley (1997); letters refer to the initial central density in units of $10^9$ g/cm$^3$: $a$: 2.0, $b$: 4.0, $c$: 5.8, $d$: 7.4, $e$: 8.2. Electron-capture supernova predictions (stars) were calculated from Figure 5 of Wanajo et al. (2013a); the numbers indicate the assumed density, relative to the default model ($\equiv 1$).

In contrast, ECSN may occur as an endpoint of evolution of a "super-AGB" star of $\sim 8-10~M_\odot$ (Nomoto 1987; Doherty et al. 2017). Such stars consist of an electron-degenerate O-Ne-Mg core, surrounded by a massive envelope. At sufficient temperature and density, electron captures may occur on $^{20}$Ne and $^{24}$Mg, leading to loss of pressure support and subsequent core-collapse and explosion. Whether such SNe exist and under what conditions is still a matter of considerable debate (see recent review by Doherty et al. 2017). In a series of papers, Wanajo et al. (Wanajo et al. 2009; Wanajo et al. 2011, 2013a, 2013b) have studied the nucleosynthesis that might occur in ECSN by starting with the evolved O-Ne-Mg core of an 8.8 $M_\odot$ star from Nomoto (1987). They found that conditions are such that n-rich isotopes including the light *r*-process (Wanajo et al. 2011), $^{48}$Ca (Wanajo et al. 2013a) and $^{60}$Fe (Wanajo et al. 2013b) may be produced in large amounts. Shown in Figure 4 are predictions of Wanajo et al. (2013a) for one set of ECSN models where each point corresponds to a model with density that was increased by up to a factor of 10, relative to the default model.

With the exception of the SNII O/Si zone, the SNIa, SNII and highest-density ECSN models all predict $^{50}$Cr depletions for $^{54}$Cr-rich ejecta (Fig. 3b), in disagreement with the grain data, if the measured anomalies at mass 50 are due to $^{50}$Cr. The SNII O/Si zone and two lowest density ECSN models do predict $^{50}$Cr enrichments, but lower than that seen for the most extreme grain, 2_37. In contrast, if the mass-50 anomalies are assumed to be due to $^{50}$Ti (Fig. 4c), all of the SN models predict $^{50}$Ti-$^{54}$Cr trends in qualitative agreement with four of the five most extreme $^{54}$Cr-rich grains. Importantly, we further note that the very low $^{48}$Ti signal measured in this grain rules out substantial incorporation of $^{48}$Ca, as this isotope is not resolvable from $^{48}$Ti under the measurement conditions we used. The inferred $^{50}$Ti/$^{48}$Ti ratio for grain 3_10 is a factor of several higher than would be expected for its $^{54}$Cr/$^{52}$Cr ratio from the SN trends (Fig. 4c) suggesting that its mass-50 anomaly may, in fact, be due to $^{50}$Cr. If so, its composition may be consistent with mixing of material from the lowest-density ECSN model (labeled 1) and more isotopically normal material (Fig. 4b).

All in all, there is broad agreement between the measured compositions of the most $^{54}$Cr-rich grains and the high-density SNIa and ECSN models shown in Figure 4, especially if some of the grains incorporated $^{50}$Ti-rich titanium. This is the first strong evidence for preserved presolar dust grains from either type of source. The quantitative agreement is best for the slow-deflagration SNIa models of Woosley (1997), but given the very limited range of model predictions available for either type of stellar explosion, we consider both types as viable candidates for the progenitors of the grains.

Although we were not able to identify the specific mineralogies of the Cr-anomalous grains, the study of Dauphas et al. (2010) pointed out that they are likely Cr-bearing spinels. For either stellar source, formation of oxide grains would require addition of O to the mix of elements produced by the explosive nucleosynthesis, most likely unburned material from the O-rich white dwarf (SNIa) or core (ECSN). Grain condensation in a high-density SNIa was modeled in an extended abstract by Yu et al. (2014). Their model produced substantial amounts of $^{54}$Cr, $^{50}$Ti, $^{48}$Ca, and other n-rich isotopes, similar to the results of Woosley (1997). They assumed the ejecta was mixed with unburned C and O, and used the equilibrium condensation code of Fedkin et al. (2010) to compute expected condensate phases at high temperature. The goal of the Yu et al. work was to test the suggestion of Dauphas et al. (2014) that correlated $^{48}$Ca and $^{50}$Ti anomalies in the solar system were carried by presolar perovskite ($CaTiO_3$) and these authors thus focused primarily on the condensation of Ti. They found that TiO condenses first, followed by perovskite ($CaTiO_3$), which consumed all the Ca. These authors did not report results for Cr-bearing phases, but Cr-bearing spinels are expected to condense at lower temperature than perovskite in a gas of solar composition (Ebel 2006). We thus postulate that the same source that produced the extreme $^{54}$Cr-rich grains first produced extremely $^{48}$Ca and $^{50}$Ti-enriched perovskite followed by the $^{54}$Cr-rich (and in some case also $^{50}$Ti-rich) oxides reported here. Identification of $^{48}$Ca- and $^{50}$Ti-rich perovskite as a presolar phase in meteorites would support this idea.

The observation that the grains with the most extreme isotopic anomalies are all very small (Fig. 2) suggests that grain condensation in the parent environments occurred rapidly. The strong apparent relationship between $^{54}$Cr/$^{52}$Cr ratio and grain size may reflect changes in gas composition during condensation within the supernova ejecta. Alternatively, it is possible that the larger grains with less isotopically anomalous Cr formed originally as more extreme small SN condensates, but subsequently grew to larger sizes by reaction with gas in the circumstellar or interstellar media, or in the protosolar disk. It is also possible that the apparent trend in Figure 2 is spurious and simply reflects two populations of grains, with the less-anomalous ones forming in different environments than the most extreme ones. We note that the NanoSIMS measurements revealed no change in isotopic composition with depth as might be expected if the grains grew from a gas whose composition was changing.

We cannot distinguish high-density SNIa from ECSN as likely sources of the anomalous grains based purely on their isotopic compositions, but we can discuss the relative merits of either source in terms of additional considerations. Firm astronomical evidence for either type of stellar explosion is lacking, let alone observations of isotopic composition or dust formation. However, we can consider expected rates of the two sources in the Galaxy and the relevant timescales of stellar evolution. Based on his nucleosynthesis results, Woosley (1997) estimated that high-density SNIa make up 2% of SNIa events in the Galaxy (~1% of all SNe). In contrast, ECSN may be much more common: with a typical initial mass function there are about half as many stars in the range of 8−10 $M_\odot$ that may produce ECSN as in the range 10−20 $M_\odot$ that end their lives as normal core-collapse SNII. Wanajo et al. (2011, 2013a) estimated that the rate of ECSN may be up to one tenth that of normal SNII, or ten times more often than high-density SNIa. Moreover, the lifetimes of stars in the 8−10 $M_\odot$ range are on order of 15−25 Myr. Because this overlaps with the lifetimes of molecular clouds, this indicates a much higher likelihood of a direct association of one or more ECSN with the protosolar cloud than that of a SNIa, whose evolutionary timescale is much longer. If, as previously proposed (e.g., Gounelle & Meynet 2012), the Sun formed in a cloud complex that underwent sequential episodes of star formation, it would be plausible for a first-generation 9−10 $M_\odot$ star to explode as an ECSN shortly before solar system formation and contribute dust that is highly enriched in n-rich isotopes. Moreover, such a scenario may explain not only the coupled anomalies in n-rich isotopes such as $^{48}$Ca, $^{50}$Ti and $^{54}$Cr, but also the recently discovered Mo isotope dichotomy in the solar system. In addition to being enriched in the n-rich isotopes we have been discussing, carbonaceous chondrite meteorites have been shown recently to also be enriched in *r*-process Mo, relative to other types of primitive meteorites (Budde et al. 2016; Kruijer et al. 2017). Wanajo et al. (2011) showed that ECSN may produce light *r*-process elements (up to ~Cd), including Mo, and this could thus provide a natural explanation for a wide range of isotopic anomalies in the solar system. One difficulty with this scenario is that prior to explosion as an ECSN, a super-AGB star would be expected to have produced copious amounts of O-rich dust with extreme $^{17}$O enrichments and $^{18}$O-depletions, due to hot-bottom burning at quite high temperature (Doherty et al. 2017) and such grains have not been seen among the presolar grain population (Nittler & Ciesla 2016). Evaluating the likelihood of molecular cloud self-pollution by an ECSN as an explanation for isotopic heterogeneities in the early solar system will likely require considerable improvements in modeling both of super-AGB stars/ECSN and the hydrodynamic evolution of molecular clouds including mixing of dust and gas and triggered star formation.

We have focused our discussion here on highly $^{54}$Cr-rich grains. The origins of the three grains with $^{54}$Cr/$^{52}$Cr ratios within error of solar but with large anomalies in other isotopes (3_24, 2_103, and 2_93; Table 1), are unclear as is the origin of $^{57}$Fe-enriched grain 2_81. Note that although this grain has a reported $^{54}$Cr/$^{52}$Cr ratio of ~1.4 times solar, the presence of an Fe-isotope anomaly indicates that the anomaly at mass 54 may in fact be due to Fe and not Cr. We hope that future multi-element measurements of these and/or similar grains will shed light on their origins.

Acknowledgements

We are grateful to Dr. Christine Floss for attempted Auger measurements on the grains. We thank NASA for supporting this work both through a Planetary Major Equipment grant to purchase the new NanoSIMS ion source and through research grant NNX17AE28G (Emerging Worlds program).

Table 1

Isotopic compositions of anomalous Orgueil grains

| Grain | diameter (nm) | $^{50}Cr/^{52}Cr$ | $^{53}Cr/^{52}Cr$ | $^{54}Cr/^{52}Cr$ | $^{50}Ti/^{48}Ti^a$ | $^{50}Ti/^{48}Ti^b$ |
|---|---|---|---|---|---|---|
| Solar | | 0.05186 | 0.11347 | 0.02821 | 0.072418 | 0.072418 |
| 2_37 | 60 | 0.317 ± 0.033 | 0.1046 ± 0.0190 | 1.6039 ± 0.1070 | 33 ± 11 | 40 ± 13 |
| 2_54a | 50 | 0.0750 ± 0.0333 | 0.1014 ± 0.0422 | 0.4117 ± 0.0898 | 0.5 ± 0.3 | 1.5 ± 0.9 |
| 2_54b | 40 | 0.0831 ± 0.0276 | 0.0886 ± 0.0327 | 0.3165 ± 0.0597 | 0.5 ± 0.1 | 1.2 ± 0.3 |
| 3_10 | 60 | 0.0806 ± 0.0125 | 0.1115 ± 0.0156 | 0.1311 ± 0.0169 | 1.9 ± 0.6 | 5.1 ± 1.5 |
| 2_50a$^c$ | 200 | 0.0764 ± 0.0181 | 0.0961 ± 0.0130 | 0.1090 ± 0.0166 | 0.080 ± 0.004 | 0.100 ± 0.004 |
| 2_90 | 65 | 0.0514 ± 0.0069 | 0.1336 ± 0.0118 | 0.0720 ± 0.0086 | | |
| 3_1 | 60 | 0.0399 ± 0.0074 | 0.0972 ± 0.0117 | 0.0696 ± 0.0097 | | |
| 2_65$^c$ | 100 | 0.0535 ± 0.0018 | 0.1157 ± 0.0028 | 0.0693 ± 0.0022 | | |
| 3_34$^c$ | 70 | 0.0499 ± 0.0057 | 0.1023 ± 0.0090 | 0.0611 ± 0.0071 | | |
| 2_116$^c$ | 200 | 0.0541 ± 0.0048 | 0.1018 ± 0.0072 | 0.0596 ± 0.0052 | | |
| 2_73 | 110 | 0.0537 ± 0.0042 | 0.1071 ± 0.0065 | 0.0586 ± 0.0049 | | |
| 2_5 | 140 | 0.0527 ± 0.0037 | 0.1024 ± 0.0057 | 0.0547 ± 0.0039 | | |
| 2_131$^c$ | 130 | 0.0491 ± 0.0031 | 0.1166 ± 0.0049 | 0.0493 ± 0.0032 | | |
| 3_5 | 130 | 0.0533 ± 0.0039 | 0.1158 ± 0.0062 | 0.0486 ± 0.0039 | | |
| 4_13 | 215 | 0.0580 ± 0.0027 | 0.1171 ± 0.0041 | 0.0386 ± 0.0023 | | |
| 2_50b$^d$ | | 0.0548 ± 0.0021 | 0.1126 ± 0.0032 | 0.0359 ± 0.0018 | | |
| 2_81$^{d,e}$ | | 0.0523 ± 0.0015 | 0.1041 ± 0.0023 | 0.0355 ± 0.0015 | | |
| 2_67 | 300 | 0.0584 ± 0.0017 | 0.1181 ± 0.0027 | 0.0355 ± 0.0014 | | |
| 4_7 | 200 | 0.0518 ± 0.0018 | 0.1131 ± 0.0029 | 0.0344 ± 0.0015 | | |
| 3_24 | 200 | 0.0401 ± 0.0026 | 0.1154 ± 0.0042 | 0.0248 ± 0.0019 | | |
| 2_103 | 100 | 0.0481 ± 0.0057 | 0.1587 ± 0.0109 | 0.0376 ± 0.0057 | | |
| 2_93 | 100 | 0.0482 ± 0.0055 | 0.0713 ± 0.0085 | 0.0243 ± 0.0049 | | |

$^a$ $^{50}Ti/^{48}Ti$ calculated with assumption that grain has solar $^{50}Cr/^{52}Cr$ ratio.
$^b$ $^{50}Ti/^{48}Ti$ calculated with assumption that grain has $^{50}Cr/^{52}Cr=0$.
$^c$ Grain is not completely resolved from one or more grains on sample mount and anomalies are thus lower limits
$^d$ Anomalous grain cannot be uniquely identified among pile of grains
$^e$ Grain also has $^{57}Fe/^{56}Fe=0.032 ± 0.002$ (1.4 ±0 .9 × solar); $^{54}Cr$ anomaly may instead be due to $^{54}Fe$